\documentclass{article}
\DeclareUnicodeCharacter{00A0}{}

\usepackage{arxiv}

\usepackage[utf8]{inputenc} 
\usepackage[T1]{fontenc}    
\usepackage{hyperref}       
\usepackage{url}            
\usepackage{booktabs}       
\usepackage{amsfonts}       
\usepackage{amsmath}
\usepackage{nicefrac}       
\usepackage{microtype}      
\usepackage{lipsum}
\usepackage{graphicx}
\usepackage{caption}
\usepackage{subcaption}
\graphicspath{ {./images/} }

\title{A Novel Approach for Earthquake Early Warning System Design using Deep Learning Techniques}

\author{
 Tonumoy Mukherjee \\
  Advanced Technology Developement Center\\
  Indian Institute of Technology Kharagpur\\
  Kharagpur 721302, India\\
  \texttt{TONUMOY@IITKGP.AC.IN} \\
   \And
 Chandrani Singh \\
  Geology \& Geophysics Department\\
  Indian Institute of Technology Kharagpur\\
  Kharagpur 721302, India \\
  \texttt{chandrani@gg.iitkgp.ac.in} \\
  \And
 Prabir Kumar Biswas \\
  Electronics \& Electrical Communication Department\\
  Indian Institute of Technology Kharagpur\\
  Kharagpur 721302, India\\
  \texttt{pkb@ece.iitkgp.ac.in} \\
}

\begin{document}
\maketitle
\begin{abstract}
Earthquake signals are non-stationary in nature and thus in real-time, it is difficult to identify and classify events based on classical approaches like peak ground displacement, peak ground velocity. Even the popular algorithm of STA/LTA requires extensive research to determine basic thresholding parameters so as to trigger an alarm. Also, many times due to human error or other unavoidable natural factors such as thunder strikes or landslides, the algorithm may end up raising a false alarm. This work focuses on detecting earthquakes by converting seismograph recorded data into corresponding audio signals for better perception and then uses popular Speech Recognition techniques of Filter bank coefficients and Mel Frequency Cepstral Coefficients (MFCC) to extract the features. These features were then used to train a Convolutional Neural Network(CNN) and a Long Short Term Memory (LSTM) network. The proposed method can overcome the above-mentioned problems and help in detecting earthquakes automatically from the waveforms without much human intervention. For the 1000Hz audio data set the CNN model showed a testing accuracy of 91.1\% for 0.2-second sample window length while the LSTM model showed 93.99\% for the same. A total of 610 sounds consisting of 310 earthquake sounds and 300 non-earthquake sounds were used to train the models. While testing, the total time required for generating the alarm was 1.68 seconds which included individual times for data collection, processing, and prediction. Taking into consideration the processing and prediction delays, the total time is thus considered to be approximately 2 seconds. This shows the effectiveness of the proposed method for EEW applications. Since the input of the method is only the waveform, it is suitable for real-time processing, thus, the models can very well be used also as an onsite earthquake early warning system requiring a minimum amount of preparation time and workload.
\end{abstract}


\section{Introduction}
\begin{figure}[htbp]
  \begin{center}
  \includegraphics[width= 6in]{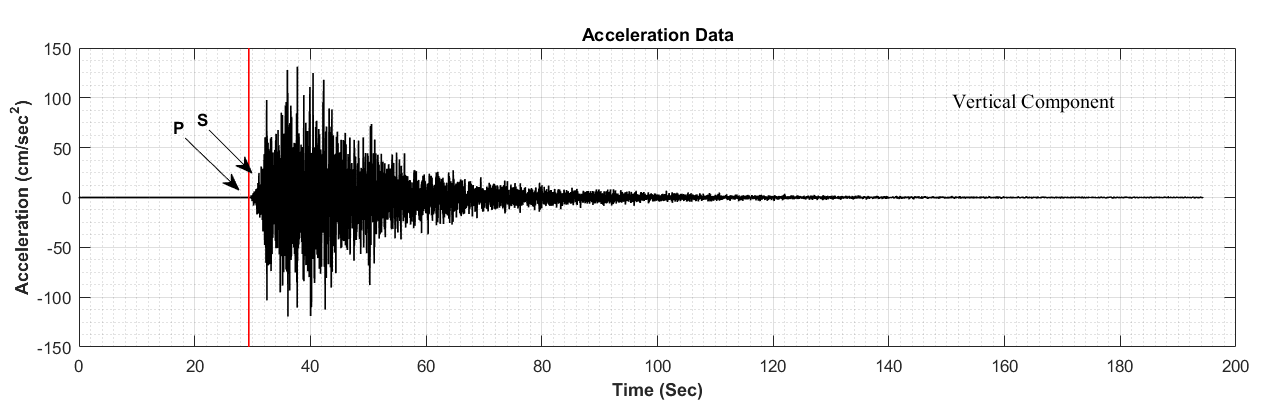}\\
 \caption{Figure showing a 6.8 magnitude earthquake acceleration signal. The red line represents the first arrival of P-wave detected by STA/LTA algorithm marked by the letter 'P' and the S-wave arrival marked by the letter 'S'.}
 \label{Figure 1}
  \end{center}
\end{figure}

Earthquakes have been an integral part of the planet earth since time immemorial. Any movement in the tectonic plates of the earth's crust releases massive amounts of energy which passes through the earth's surface as seismic waves and results in mild to tremendous shaking. This whole process although seems to be very long but happens within a few minutes. As a result, the time to respond to these occurrences have increased many folds. Various scales of 1-10 have been adopted as the magnitude indicator of earthquakes. For addressing a wider range of earthquake sizes, the moment magnitude scale, abbreviated $M_W$, is preferred and is applicable globally \cite{hanks1979moment}. Magnitudes of earthquakes are exponential. To put it simply, for each whole number that we go up on a magnitude scale, the amplitude of the ground motion goes up by a factor of 10 when recorded by a seismograph \cite{kanamori1978quantification}. Thus, by using this scale as a reference, it is realized that the level of ground shaking caused by a magnitude 2 earthquake would be ten times more than a magnitude 1 earthquake (and 32 times as much energy would be released). To put that into context, if a magnitude 1 earthquake releases as much energy as blowing up 6 ounces of TNT, a magnitude 8 earthquake would release as much energy as detonating 6 million tons of TNT. Major earthquakes can cause significant damage to life as well as property. Preventing an earthquake from occurring is an illogical thing to do. We need to focus on how to mitigate the devastation caused by these events. For that,  we need some robust and reliable prediction methods. There were a few successful predictions. The Haicheng earthquake(1975) is a perfect example of a successful prediction. Precursors for this event included a foreshock sequence, peculiarity in animal behavior, and anomalies like geodetic deformation, groundwater level differences\cite{precursor}. But these abnormalities were not present in other major earthquakes. Hence there is no generic precursor for earthquakes. This is where modern approaches of machine learning and deep learning comes into play. Although researches in seismology using machine learning and deep learning are quite limited, with the huge amount of data that are accessible to researchers, good researches like foreshock identification in real-time using deep learning \cite{Vikraman2016a}, Earthquake detection and location using convolutional neural network \cite{Perol2018a} and Machine Learning Seismic Wave Discrimination \cite{Li2018a} have been done.  Thus having an idea about the occurrence of an earthquake as early as possible will result in developing a certain amount of alertness to respond to such events with ease. Unlike various methods of detecting an earthquake such as STA/LTA (Short-Term Average/Long-Term Average) algorithm as shown in Figure: \ref{Figure 1} \cite{wu2007}, another possible way of detecting and classifying earthquakes using their sounds, have been explored in this research. Sound is an effect produced by anything physical including earthquakes \cite{sound1} and is one of the most common effects reported during or immediately after the felt tremors caused by them\cite{sound2}. An earthquake of magnitude 5, will have a sound or vibration different from another earthquake whose magnitude is 4 \cite{eq-snd}. Also, similar magnitude earthquakes should have similar vibrations or sounds that are indistinguishable by us but can be picked up very efficiently by neural networks.\\ 
In this paper, two types of machine learning models were used to train a system with data consisting of earthquake and non-earthquake sounds. Convolutional Neural Network (CNN) and Long Short Term Memory (LSTM) models were used for the purpose. The former model showed a testing accuracy of 91.102\% for a 0.2-second audio data sample and the latter showed a testing accuracy of 93.999\% for the same.

\section{Deep Learning and Seismology}
With the rapid increase in the seismic data quantity, major challenges are faced by modern seismology in the fields of data analyzing and processing techniques. Most of the many popular techniques that are used in major data centers date back to the time when the amount of seismic data was small and the computational power was limited.\\
Today with the advancements in the fields of machine learning and deep learning, scientists and researchers can very easily extract useful information from voluminous data as they provide a large collection of tools to work with. Once trained with sufficient data, deep learning models just like humans, can acquire their knowledge by extracting features from raw data \cite{goodfellow2016deep} to recognize natural objects and make expert-level decisions in various disciplines. Besides, the high computational costs for training such networks are balanced by their low-cost online operation \cite{Perol2018a}. Advantages like these, make deep learning suitable for applications in real-time seismology and earthquake early warning (EEW).\\
This paper would look into two commonly used and widely known Deep learning models, CNN and LSTM respectively and the accuracy achieved by them in detecting and classifying earthquakes from their sounds. The architecture of both the networks would be looked at in detail in the following sections. 

\subsection{Convolutional Neural Network (CNN) Architecture}
\begin{figure}[!tbp]
  \begin{subfigure}[b]{0.81\textwidth}
    \includegraphics[width=\textwidth]{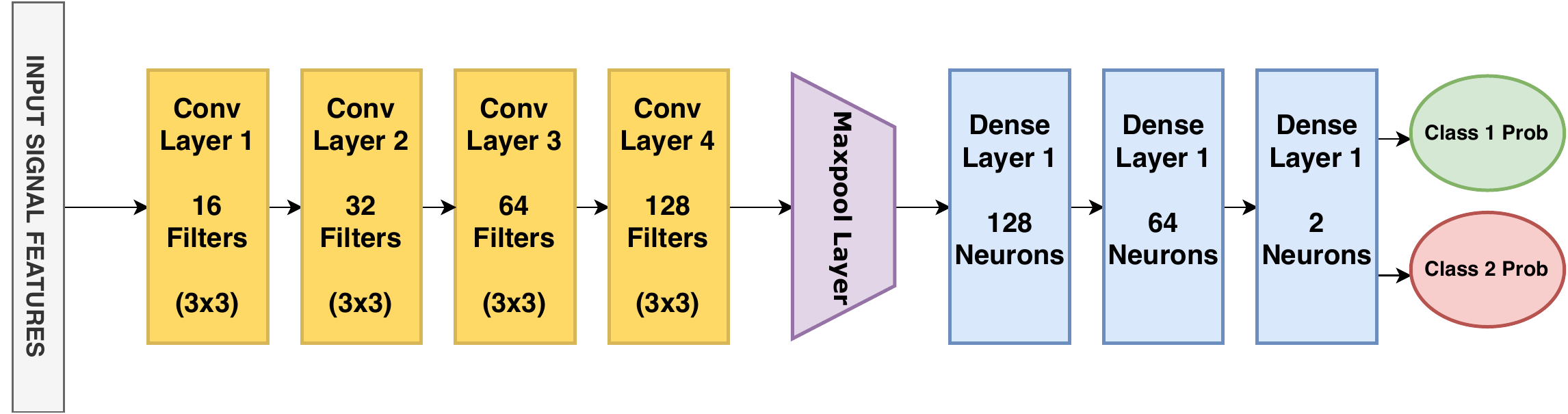}
    \caption{Block Diagram of the CNN Model used}
    \label{Figure 2(a)}
  \end{subfigure}
  \hfill
  \begin{subfigure}[b]{0.18\textwidth}
    \includegraphics[width=\textwidth]{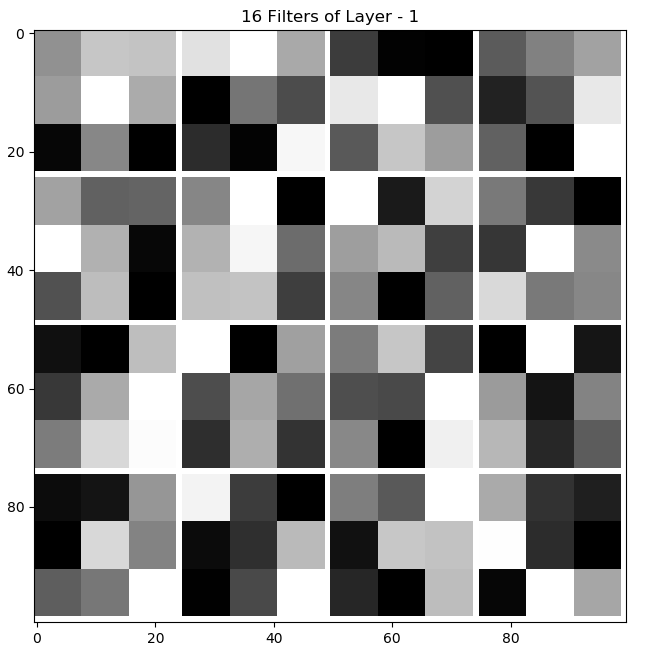}
    \caption{16 Filters of the First Convolutional Layer arranged in a 4X4 Matrix}
    \label{Figure 2(b)}
  \end{subfigure}
  \caption{CNN Block Diagram with First Convolutional Layer Filters}
\end{figure}

The proposed architecture for the CNN  model has 4 convolutional layers, 1 maxpool layer, and 3 dense layers. \ref{Figure 2(b)} shows the first layer of the 4 convolutional layers consisting of 16 filters built with 3x3 convolution with 'Relu' as the activation function and 1x1 stride. All the parameters for the second, third, and fourth layers remain the same except for the number of filters in each layer multiplies two times with the number of filters in the previous layer. Or in other words, 16 filters in the first layer, 32 filters in the second layer, 64 in the third and 128 in the final layer (shown in \ref{Figure 2(a)}).

The idea behind increasing the number of filters in each layer is to be more specific about the features as the data starts to convolve through each layer. A kernel of 2x2 has been used for the maxpool layer. The three dense layers after the maxpool layer consist of 128, 64, and 2 neurons so as to pool down the features for the final 2 class classification. The first two dense layers use 'Relu' as their activation function whereas the last dense layer uses 'Softmax' as its activation function as we use categorical cross-entropy for multi-class classification purposes and 'adam' as the optimizer. 

\subsection{Long Short Term Memory (LSTM) Architecture }
\begin{figure}[htbp]
  \begin{center} 
  \includegraphics[width=0.8\textwidth]{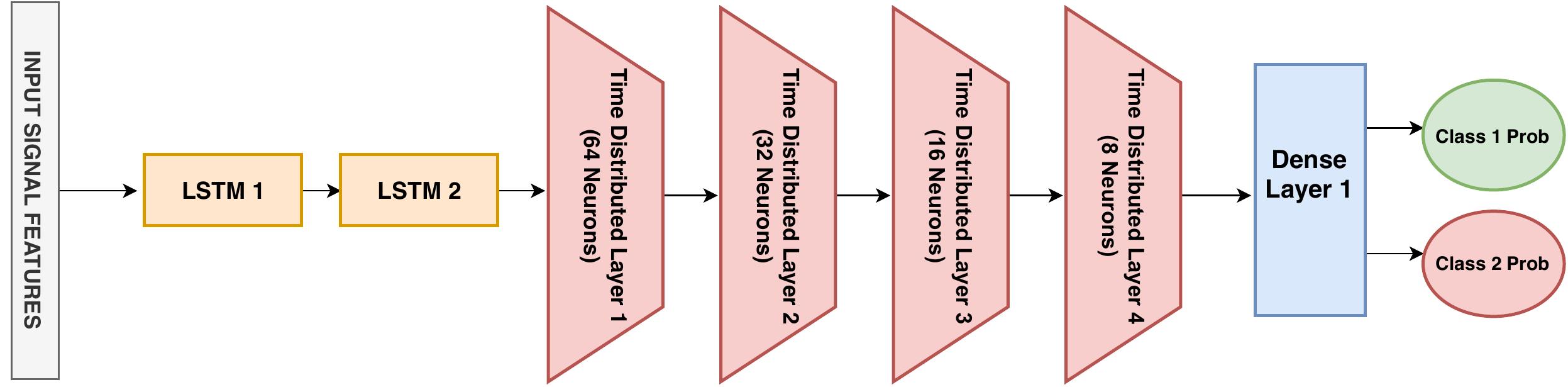}\\
 \caption{Block Diagram of the LSTM Model used}
 \label{Figure 3}
  \end{center}
\end{figure}

The proposed architecture for the LSTM model (shown in Figure: \ref{Figure 3}) has 2 LSTM layers consisting of 128 neurons each. 4 time distributed fully connected layers of 64, 32, 16, and 8 neurons respectively are added after the 2 LSTM layers with 'relu' as their activation function. Lastly, a dense layer consisting of 2 neurons is added for the final 2 class classification with 'softmax' as its activation function and 'adam' as the optimizer. 

\section{Data}
Indian Earthquake acceleration data of the past 10 years and magnitudes approximately between 2-8$M_W$ was collected from PESMOS (Program for Excellence in Strong Motion Studies, IIT Roorkee) \cite{mittal2012indian} and EGMB (Eastern Ghats Mobile Belt) Data of earthquakes with magnitudes ranging between 2-5$M_W$ Collected at Geology and Geophysics Lab IIT Kharagpur, were also used. 

\begin{table*}[th]
\caption{Training Data for Deep Learning Models (CNN \& LSTM)}
\label{tab:train}
\resizebox{\textwidth}{!}{%
\begin{tabular}{|c|c|c|c|c|}
\hline
\textbf{Data Source} & \textbf{Data Type \& Format} & \textbf{Feature Vector Dimensions} & \textbf{No. of Earthquake  Data} & \textbf{No. of Non-Earthquake  Data} \\ \hline
PESMOS (IIT Roorkee) & Audio,  .wav                 & 9 x 13                             & 212                              & 0                                    \\ \hline
EGMB (IIT Kharagpur) & Audio, .wav                  & 9 x 13                             & 68                               & 280                                  \\ \hline
\end{tabular}%
}
\end{table*}

\begin{table*}[th]
\caption{Testing Data for Deep Learning Models (CNN \& LSTM)}
\label{tab:test}
\resizebox{\textwidth}{!}{%
\begin{tabular}{|c|c|c|c|c|}
\hline
\textbf{Data Source} & \textbf{Data Type \& Format} & \textbf{Feature Vector Dimensions} & \textbf{No. of Earthquake  Data} & \textbf{No. of Non-Earthquake  Data} \\ \hline
PESMOS (IIT Roorkee) & Audio,  .wav                 & 9 x 13                             & 11                               & 0                                    \\ \hline
EGMB (IIT Kharagpur) & Audio, .wav                  & 9 x 13                             & 21                               & 18                                   \\ \hline
\end{tabular}%
}
\end{table*}

Table \ref{tab:train} and Table \ref{tab:test} shows the Train-Test Data split for both the Deep Learning models. The collected dataset was replicated twice. The first copy of the dataset was converted into corresponding audio signals of .wav format by keeping the original sensor sampling rate 200Hz and the second copy of the same dataset was converted into corresponding audio signals of the same format whose sampling rate was increased to 1000Hz using upsampling techniques. The sampling rate of the upsampled audio signals was estimated using trial and error method so as to hear and clearly notice the change of the signal with respect to time. A total of 610 sounds consisting of 310 earthquake sounds and 300 non-earthquake sounds were used to train the models. Both the audio signal datasets of 200Hz and 1000Hz were fed as inputs to both the above models.

\section{Methodology}
\subsection{Data Preparation }
Before training the models, proper cleaning of the audio data was done and essential features from the data were then extracted using popular speech processing methodologies of Filter Banks and Mel-Frequency Cepstral Coefficients (MFCCs).

\subsection{Pre-Emphasis}
The very first step of data processing included filtering the data using a Pre-Emphasis filter. This was mainly done to amplify the high frequencies. Apart from amplification, the filter helped in balancing the frequency spectrum since higher frequencies usually have smaller magnitudes compared to lower frequencies. The filter also was able to improve the Signal-to-Noise Ratio (SNR).
The first order filter is represented by the equation  :
\begin{equation}
y(t) = x(t)-\alpha x(t-1)    
\end{equation}
 was used to apply the pre-emphasis filter over the range of audio signal data.
From the typical values of 0.95 and 0.97, the former was used for the filter coefficient $\alpha$.

\subsection{Framing and FFT}
\label{sec:Framing}
After pre-emphasis filtering, the data was divided into short-time frames to avoid the loss of frequency contours of the signal over time for performing Fourier transform. A good approximation of the frequency contours of the signal was achieved by concatenating adjacent frames and applying the Fourier transform over those short-time frames. Popular settings of 25ms for the frame-size and 10ms stride were used for framing the data. A Hamming window function (shown in Eq.2) was applied after the signal was sliced into frames. 

\begin{equation}
w[n] = 0.54-0.46\cos\frac{2\pi n}{N-1},   \text{for}\ 0 \leq n \leq N-1 
\end{equation}
where N represents the number of frames in which the signal was divided. 
\\After dividing the signal into frames, an N-point FFT was performed on each frame to calculate the frequency spectrum, which also happens to be the Short-Time Fourier-Transform (STFT), where N is typically 256 or 512 (256 in this case).

\subsection{Filter Banks and Mel-Frequency Cepstral Coefficients (MFCCs)}
\label{sec:Filter}

\begin{figure}[h]
\includegraphics[width=6in]{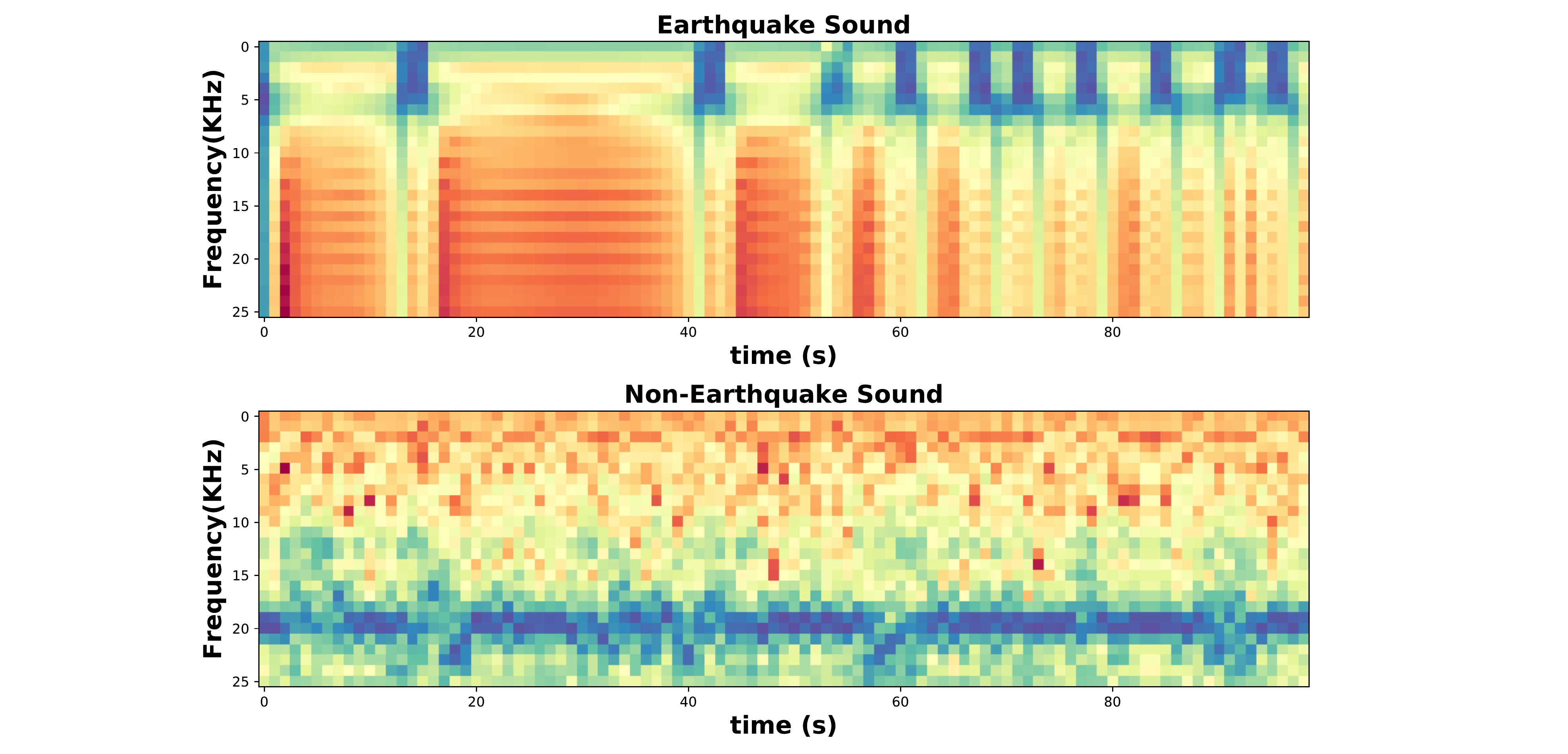} 
\caption{Features extracted from the signal for training the models after applying the filter bank to the power spectrums of the short-time frames of the signal.}
\label{Figure 4}
\end{figure}

The rationale behind using filter banks was to separate the input signal into its multiple components such that each of those components carries a single frequency sub-band of the original signal. Triangular filters, typically 26, were applied on a Mel-scale to the power spectrum of the short-time frames to extract the frequency bands as shown in Figure: \ref{Figure 4}.

The formula for converting from frequency to Mel scale is given by:
\begin{equation}
    M(f) = 1125\times ln(1+\frac{f}{700})    
\end{equation}
To go back from Mels to frequency, the formula used is given by:
\begin{equation}
    M^{-1}(m) = 700 \times(e^{m/1125}-1)
\end{equation}

\begin{figure}[h]
\includegraphics[width=6in]{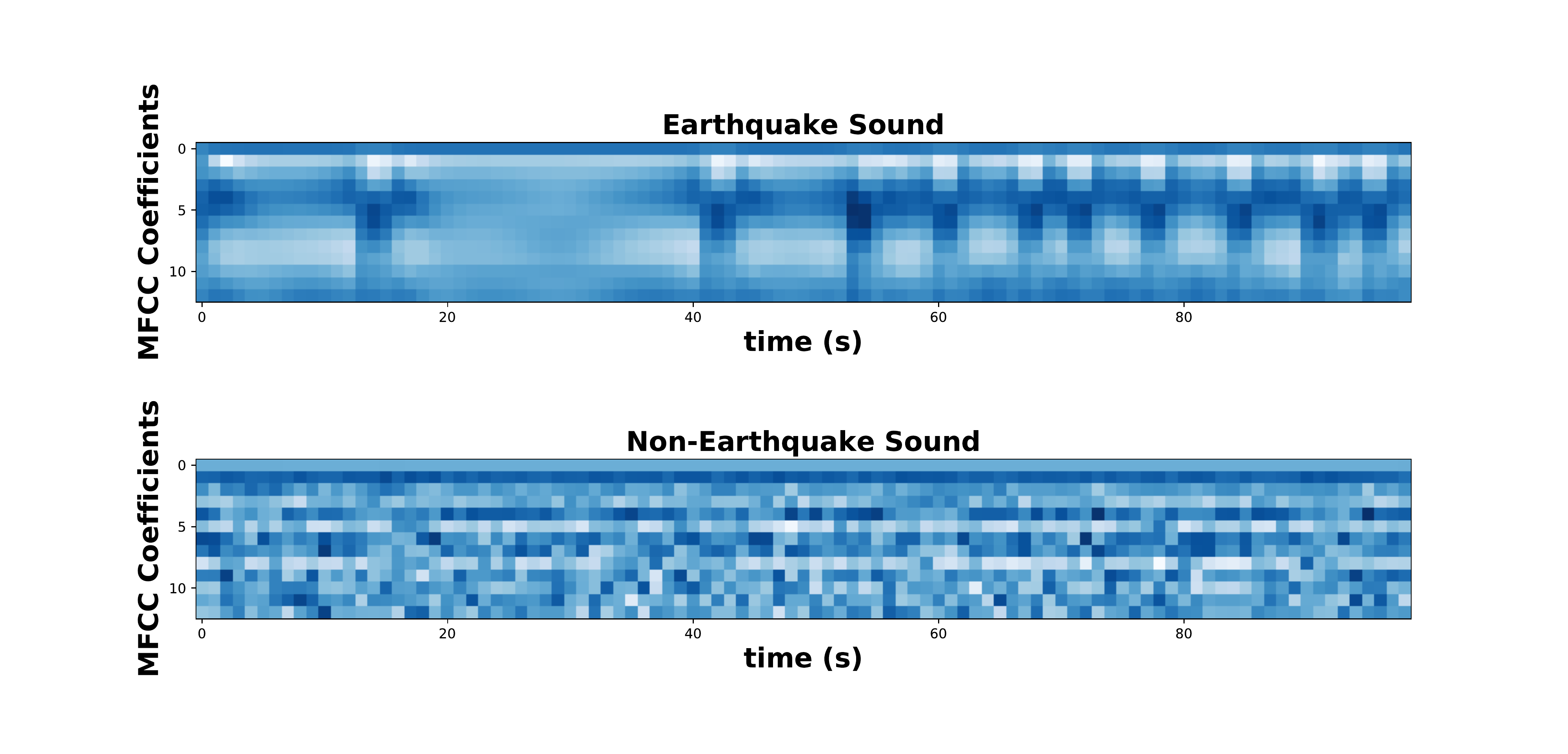}
\caption{Mel-Frequency Cepstral Coefficients (MFCCs) as extracted from the short frames in which the signal was divided.}
\label{figure 5}
\end{figure}

MFCC is a biologically inspired and by far the most successful and most used feature in the area of speech processing \cite{hossan2010novel}. The algorithm was used in volcano classification also \cite{volcano}. For speech signals, the mean and the variance changes continuously with time, and thus it makes the signal non-stationary \cite{frei2006intrinsic}. Similarly like speech, earthquake signals are also non-stationary \cite{hammond1996analysis} as each of them has different arrival of P, S, and surface waves. Therefore, normal signal processing techniques like Fourier transform, cannot be directly applied to it. But, if the signal is observed in a very small duration window (say 25ms ), the frequency content in that small duration appears to be more or less stationary. This opened up the possibility of short-time processing of the earthquake sound signals. The small duration window is called a frame, discussed in section \ref{sec:Framing}. For processing the whole sound segment, the window was moved from the beginning to the end of the segment consistently with equal steps, called shift or stride. Based on the frame-size and frame-stride, it gave us M frames. Now, for each of the frames as depicted in Figure:  \ref{figure 5}, MFCC coefficients were computed.

\begin{figure}[th]
  \centering
  \includegraphics[width=5in]{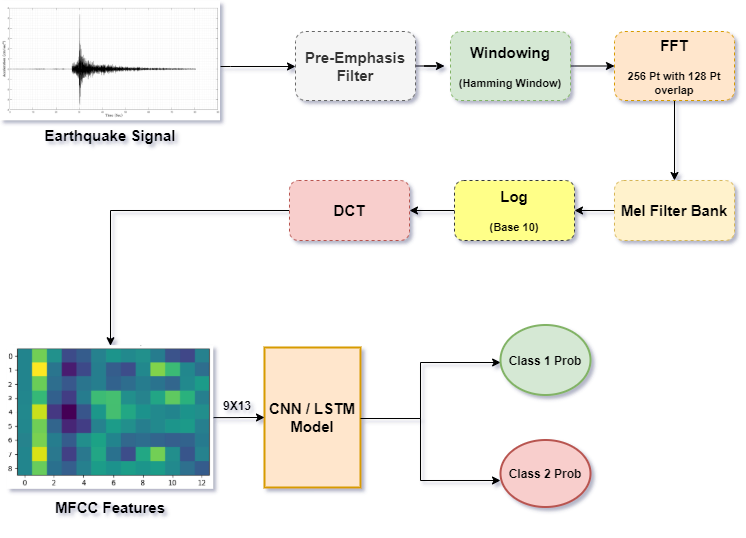}\\
  \caption{Block diagram of the entire System}
  \label{Figure 6}
\end{figure}

Moreover, the filter bank energies computed were highly correlated since all the filterbanks were overlapping. This becomes a problem for most of the machine learning algorithms. To reduce autocorrelation between the filterbank coefficients and get a compressed representation of the filter banks, a Discrete Cosine Transform (DCT) was applied to the filterbank energies. This also allowed the  use of diagonal covariance matrices to model the features for training. Also, 13 of the 26 DCT coefficients were kept and the rest were discarded due to the fact that fast changes in the filterbank energies are represented by higher DCT coefficients. These fast changes resulted in degrading the model performances. Thus a small improvement was observed by dropping them.\\
The overall system representation could be better understood from Figure: \ref{Figure 6}.

\section{Results and Discussion}
\begin{figure}[htbp]
 \centering
  \includegraphics[width=6in]{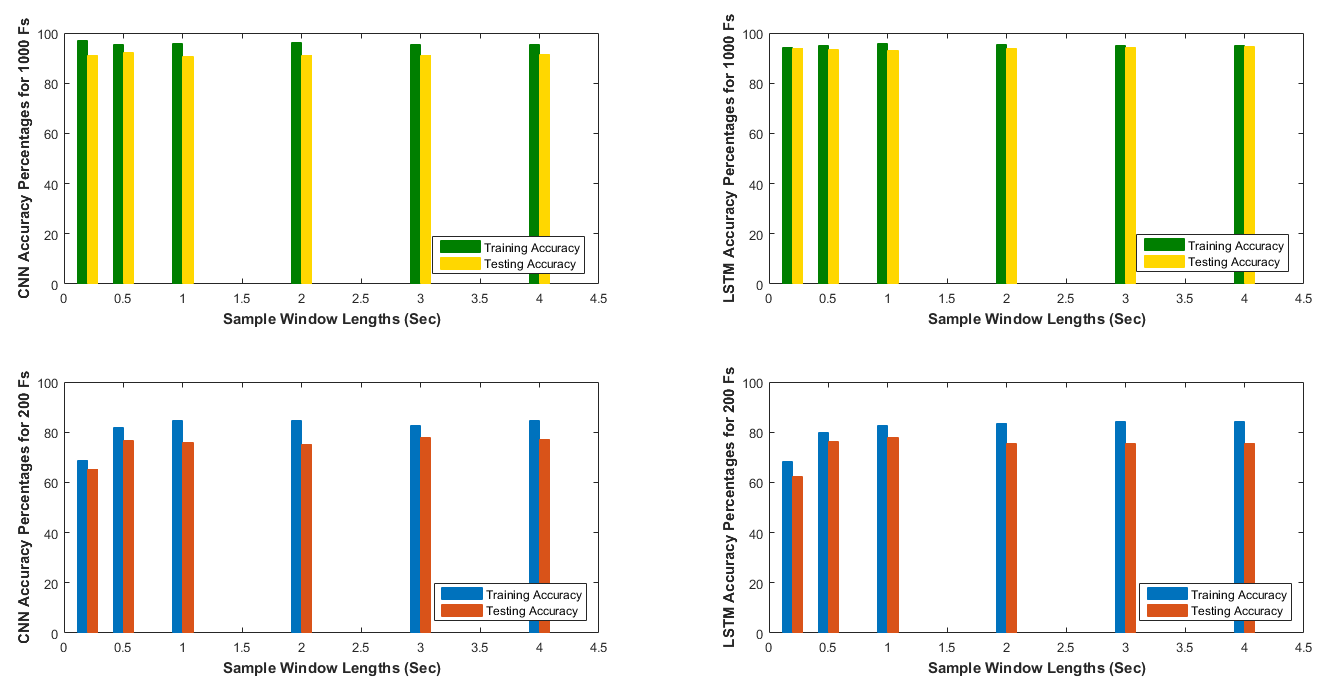}\\
  \caption{Testing and Training Accuracy comparisons of CNN and LSTM Models over several sampling window time frames for the audio signal data sets of 1000Hz \& 200Hz sampling frequencies}
  \label{Figure 7}
\end{figure}

The CNN and the LSTM models performed almost similarly for the 200Hz audio data set, but significant improvements in the train-test accuracy percentages are observed for the 1000Hz data set. For 1000Hz audio data set the CNN model showed a testing accuracy of 91.102\% for 0.2-second sample window length while the LSTM model showed 93.999\% for the same (shown in Figure: \ref{Figure 7}). This observation can be backed by the fact that LSTMs performs better for sequential or time-series data classifications \cite{LSTM}. 

\begin{figure}[htbp]
  \centering
  \includegraphics[width=4in]{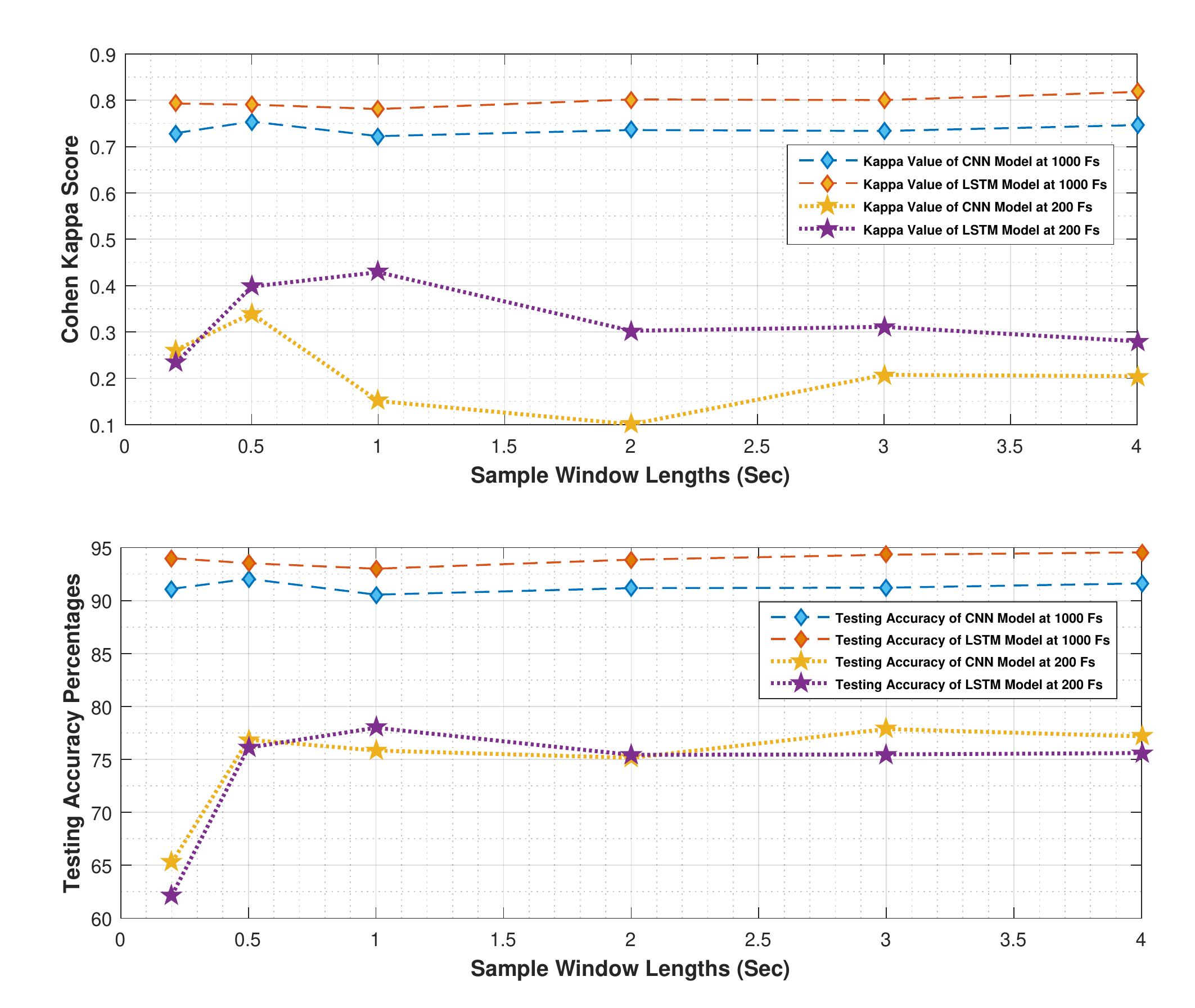}\\
  \caption{Testing Accuracy and Cohen Kappa Score comparisons of the CNN and the LSTM Models over several sampling window time frames for the audio signal data sets of 1000Hz and 200Hz sampling frequencies}
  \label{Figure 8}
\end{figure}

\begin{figure}[htbp]
  \centering
  \includegraphics[width=4in]{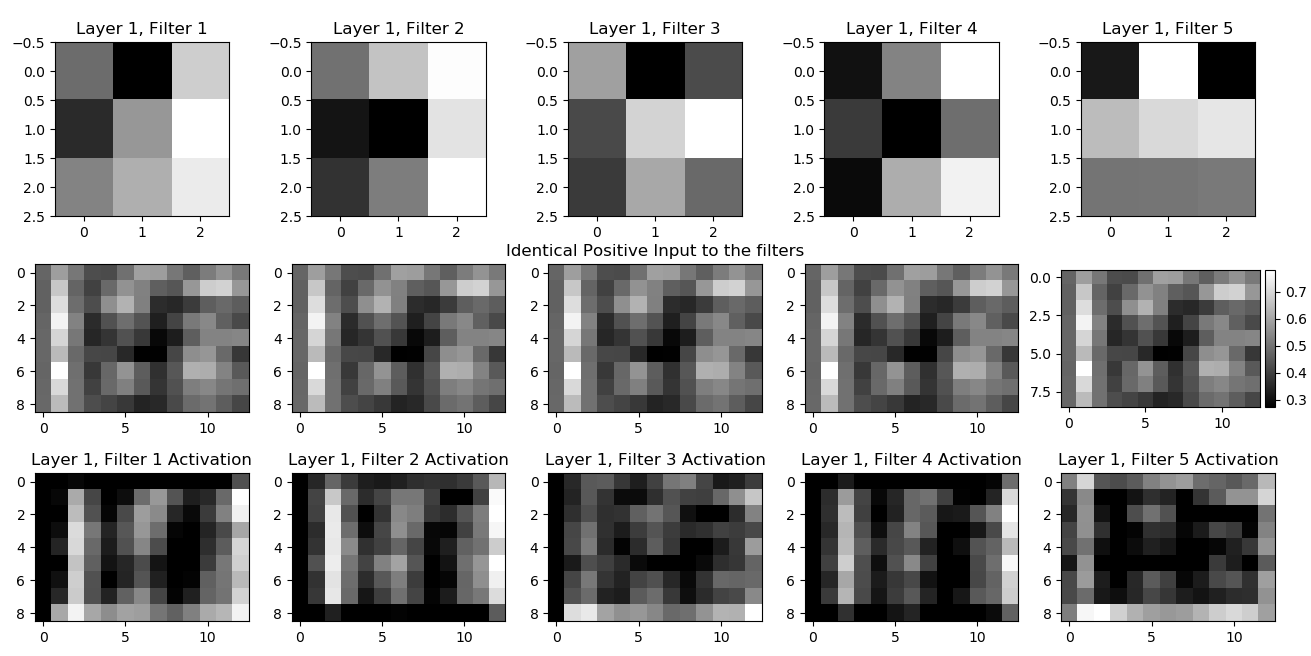}\\
  \caption{1st Convolutional layer activations for Earthquake Data with corresponding Filters}
  \label{Figure 9}
\end{figure}

\begin{figure}[htbp]
  \centering
  \includegraphics[width=4in]{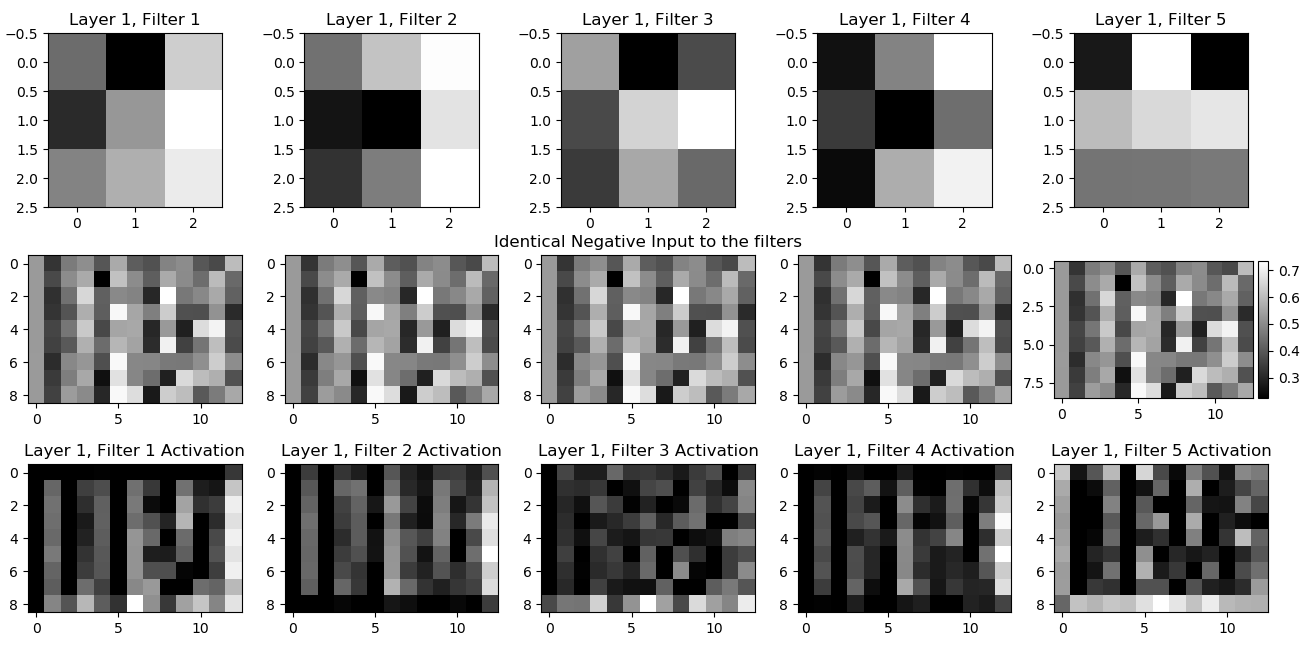}\\
  \caption{1st Convolutional layer activations for Non- Earthquake Data with the same Filters as used for Earthquake data}
  \label{Figure 10}
\end{figure}

The Kappa statistics(values), generally used for comparing an Observed Accuracy with an Expected Accuracy (random chance), was used for validating the model accuracies for both the data sets (shown in Figure: \ref{Figure 8}).\\
For both data classes, activations by random 5 out of 16 filters of the first layer of the CNN Model along with their inputs is represented by Figure: \ref{Figure 9} and Figure: \ref{Figure 10} respectively.

The time required for generating the alarm by the CNN model includes 1.28 seconds to gather enough data to start the MFCC computations, 8ms for processing and a prediction time of 0.2 seconds. The summation gives a result of 1.68 seconds. Taking into consideration the processing and prediction delays, the total time for the CNN model is thus considered to be $\simeq{2}$seconds.

LSTMs being computationally a bit more expensive, the processing and the prediction times were 10ms and 0.5 seconds respectively, giving a total of 2.19 seconds. Taking into consideration the processing and prediction delays, the total time for the LSTM model is thus considered to be $\simeq{2.5}$seconds.

\begin{table*}[th]
\caption{Overall Comparision between standard STA/LTA algorithm and the proposed algorithms}
\label{tab:comparision}
\resizebox{\textwidth}{!}{%
\begin{tabular}{|c|c|c|c|c|c|}
\hline
\textbf{Algorithm} &
  \textbf{Data Source} &
  \textbf{Data Sampling Frequency} &
  \textbf{Time to Alarm} &
  \textbf{Accuracy} &
  \textbf{Prerequisites} \\ \hline
STA/LTA &
  \begin{tabular}[c]{@{}c@{}}PESMOS (IIT Roorkee)\\ \&\\ EGMB (IIT Kharagpur)\end{tabular} &
  200 Hz &
  3 Seconds after P-wave arrival&
  \begin{tabular}[c]{@{}c@{}} 95.43\%\\ (Heavily dependant on Prerequisites)\end{tabular} &
  \multicolumn{1}{l|}{\begin{tabular}[c]{@{}l@{}}1. Proper user-defined Thresholds\\ 2. Different Thresholds for different Regions\\ 3. Threshold kept high for Strong Motion Events \\     (more earthquakes missed, lesser false alarms)\\ 4. Threshold kept low for Weak Motion Events\\     (fewer earthquakes missed, more false alarms)\end{tabular}} \\ \hline
\begin{tabular}[c]{@{}c@{}}MFCC with CNN Model\\ (Proposed Method)\end{tabular} &
  \begin{tabular}[c]{@{}c@{}}PESMOS (IIT Roorkee)\\ \&\\ EGMB (IIT Kharagpur)\end{tabular} &
  1000 Hz &
  2 Seconds after P-wave arrival &
  91.1\% &
  None \\ \hline
\begin{tabular}[c]{@{}c@{}}MFCC with LSTM Model\\ (Proposed Method)\end{tabular} &
  \begin{tabular}[c]{@{}c@{}}PESMOS (IIT Roorkee)\\ \&\\ EGMB (IIT Kharagpur)\end{tabular} &
  1000 Hz &
  2.5 Seconds after P-wave arrival &
  93.99\% &
  None \\ \hline
\end{tabular}%
}
\end{table*}

\begin{figure}[th]
  \centering
  \includegraphics[width=6in]{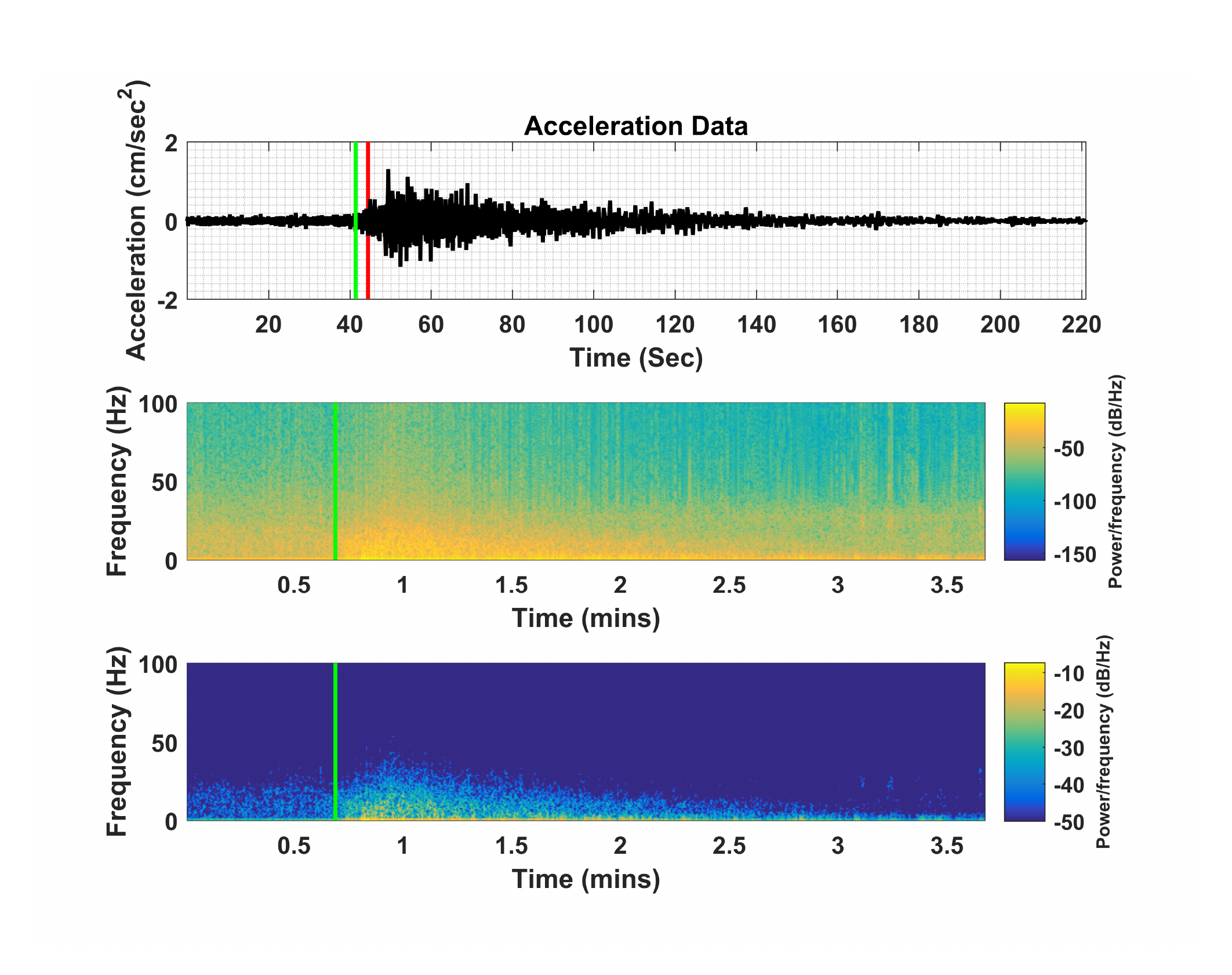}\\
 \caption{Figure Showing 3 subplots where the first plot shows incorrect P-wave arrival detection by STA/LTA, marked by the red line, due to poor threshold parameters, leading to late detection of P-wave and in-turn leading to late alarm generation. The second plot shows the spectrogram with almost appropriate detection of the P-wave first arrival marked by the green line and the third plot shows the same spectrogram with a -50dB threshold applied, so as to get a clear view of the first P-wave arrival.}
 \label{Figure 11}
\end{figure}

It was also observed that by increasing the sampling rate of the audio signals, the training and the testing accuracies of both CNN and LSTM model increased. 
This is one of the very first applications of Machine learning in the field of earthquake detection using MFCC's and Filterbank Coefficients which are generally used in the field of Speech Recognition. An earthquake is basically understood well by three types of waves namely P-wave, S-wave, and surface wave. Interaction of these waves with the surrounding medium gives an earthquake its intensity. Any wave is a vibration. Any vibration has some sound associated with it. It might be inaudible to the human ears, but the sound remains. In real-time, it is difficult to identify and classify events based on classical approaches like peak ground displacement, peak ground velocity, or even the widely recognized algorithm of STA/LTA as they require extensive research to determine basic thresholding parameters so as to trigger an alarm. Many times due to human error or other unavoidable natural factors such as thunder strikes or landslides, the conventional algorithms may end up raising a false alarm (shown in Figure: \ref{Figure 11}).

Table \ref{tab:comparision} shows a detailed comparision between the standard STA/LTA algorithm and the proposed Deep Learning models. The main disadvantage of STA/LTA algorithm is that it has to be tweaked differently for different types of event detections. The user defined threshold which acts as the trigger, varies from region to region due to its direct dependence on the geographical and topological features of a particular region. Also, the threshold is kept high for strong motion events and low for weak motion events. This results in serious ambiguity as both the features can't be used together. If the threshold is high, more earthquakes are missed but lesser false alarms are generated and if the threshold is low, less earthquakes are missed but more false alarms generated. The proposed method can overcome these problems as it can extract essential features of a raw sound or vibrational data without manually hand engineering them and doesn’t require any knowledge or expertise in the relevant field. It is also invariant to small variations in occurrence in time or position and can understand representations of data with multiple levels of abstraction. Since the input of the method is only the waveform, it is suitable for real-time processing, thus, the models can very well be used also as an onsite earthquake early warning system requiring a minimum amount of preparation time and workload. Until now, the use of earthquake early warning systems for earthquake disasters are mainly limited by false alarm generation and delay in detection. By using the suggested approach, these problems can be overcome, leading to automatic, fast, and accurate detection of earthquake seismic signals.

\subsection{Seismic Sensor Prototype Hardware Design and Experimental Setup for Model Testing}
\begin{figure}[ht]
  \begin{center} 
  \includegraphics[width = \textwidth]{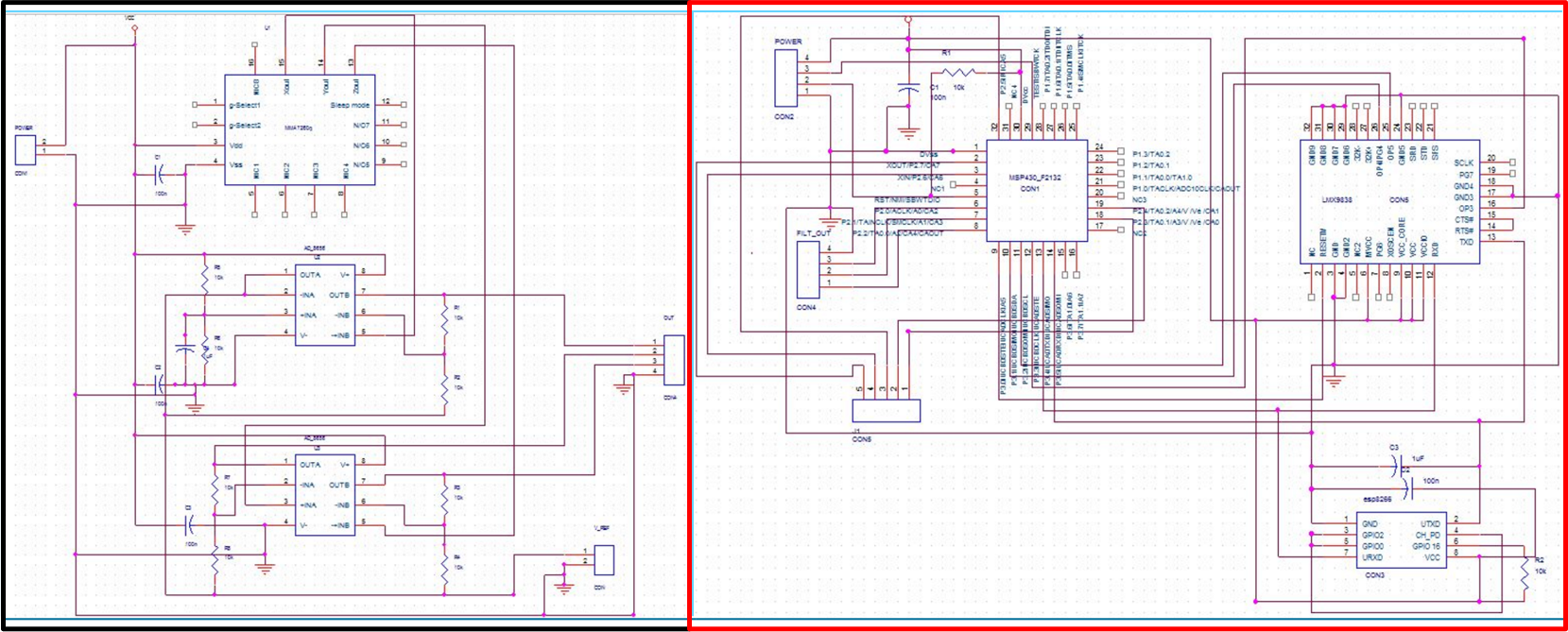}\\
 \caption{Schematic of an accelerometer with front-end amplifiers (box in black), and a microcontroller 
with wireless transceiver (box in red).}
 \label{Figure 12}
  \end{center}
\end{figure}

\begin{figure}[htbp]
  \centering
  \begin{subfigure}{0.7\textwidth}
    \includegraphics[width=1\textwidth]{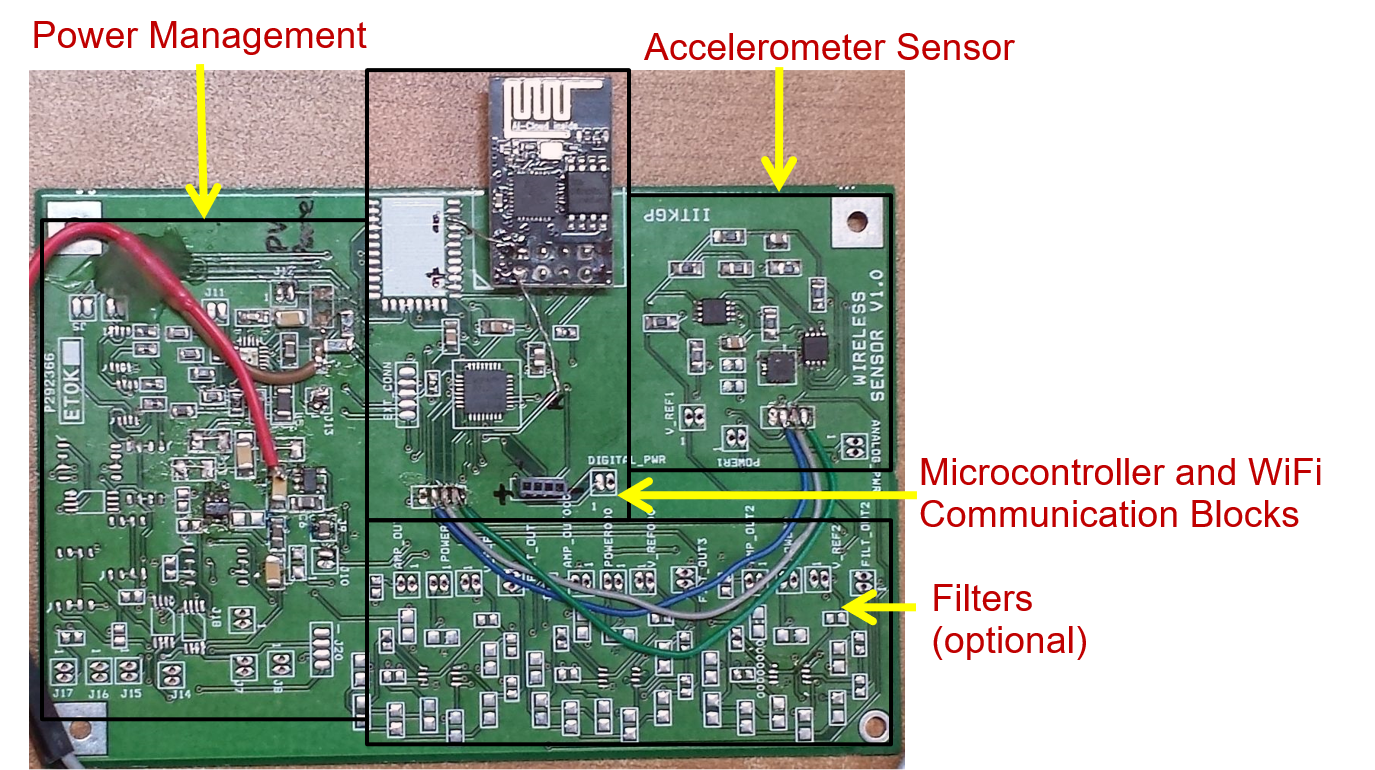}
    \caption{Photo of the prototype circuit board for implementing a wireless seismic sensor node.}
    \label{Figure 13(a)}
  \end{subfigure}
   \begin{subfigure}{0.4\textwidth}
    \includegraphics[width=1\textwidth]{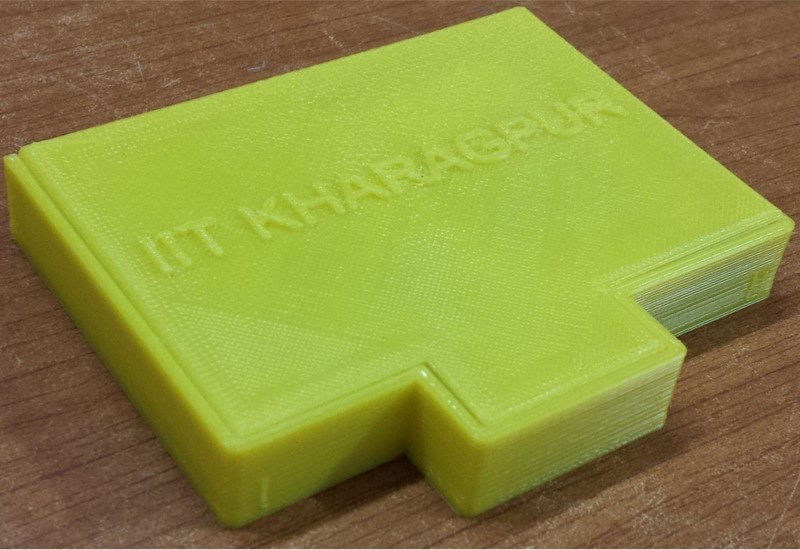}
    \caption{A 3D printed packaging scheme for the wireless seismic sensor prototype.}
    \label{Figure 13(b)}
  \end{subfigure}
  \begin{subfigure}{0.9\textwidth}
    \includegraphics[width=0.6\textwidth]{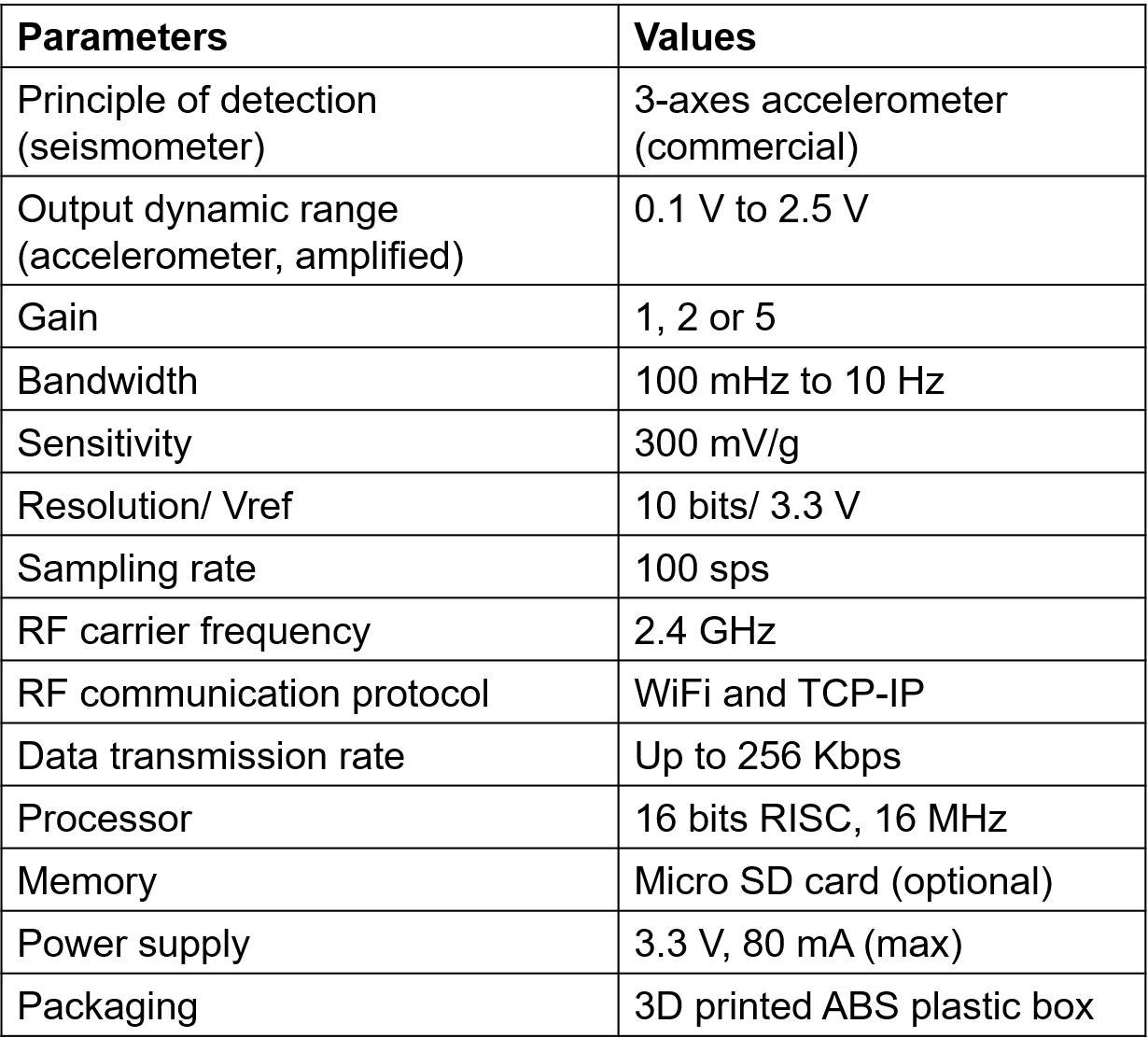}
    \caption{Specifications of the Seismic sensor Prototype.}
    \label{Figure 13(c)}
  \end{subfigure}
  \caption{Prototype Sensor PCB with its 3D printed packaging and Specifications}
\end{figure}

The main reason for developing the sensor hardware prototype was to test the validity and robustness of the proposed models. Also, it was very crucial to understand and evaluate the complexities and solve the challenges at a system level as it is targeted to deploy the system in earthquake prone areas. It was also essential for comparing the system results with already present commercial solutions.\\   
The hardware is developed with the help of commercial equipments to mimic the vibrations from an actual earthquake. Figure: \ref{Figure 12} depicts a schematic of the prototype sensor hardware used. Figure: \ref{Figure 13(a)} shows the actual PCB with boxes representing the utility of sections. On the hard left we have the power management block to supply a steady source of power to the circuit (3.3V,80mA). On the top right we have the accelerometer sensor acting as the principle of detection (seismometer) having an output dynamic range of 0.1V to 2.5V and a sensitivity of 300mV/g. In between the Power Management and the accelerometer sensor blocks we have the microcontroller and wifi communication blocks to receive and transmit the sensor data wirelessly. The Wifi has a bandwidth of 100mHz to 10Hz, RF carrier frequency of 2.4GHz with a data transmission rate upto 256Kbps. The processor is of a 16 bit RISC architecture with an operating frequency of 16MHz. 

\begin{figure}[ht]
  \begin{center} 
  \includegraphics[width = 6in]{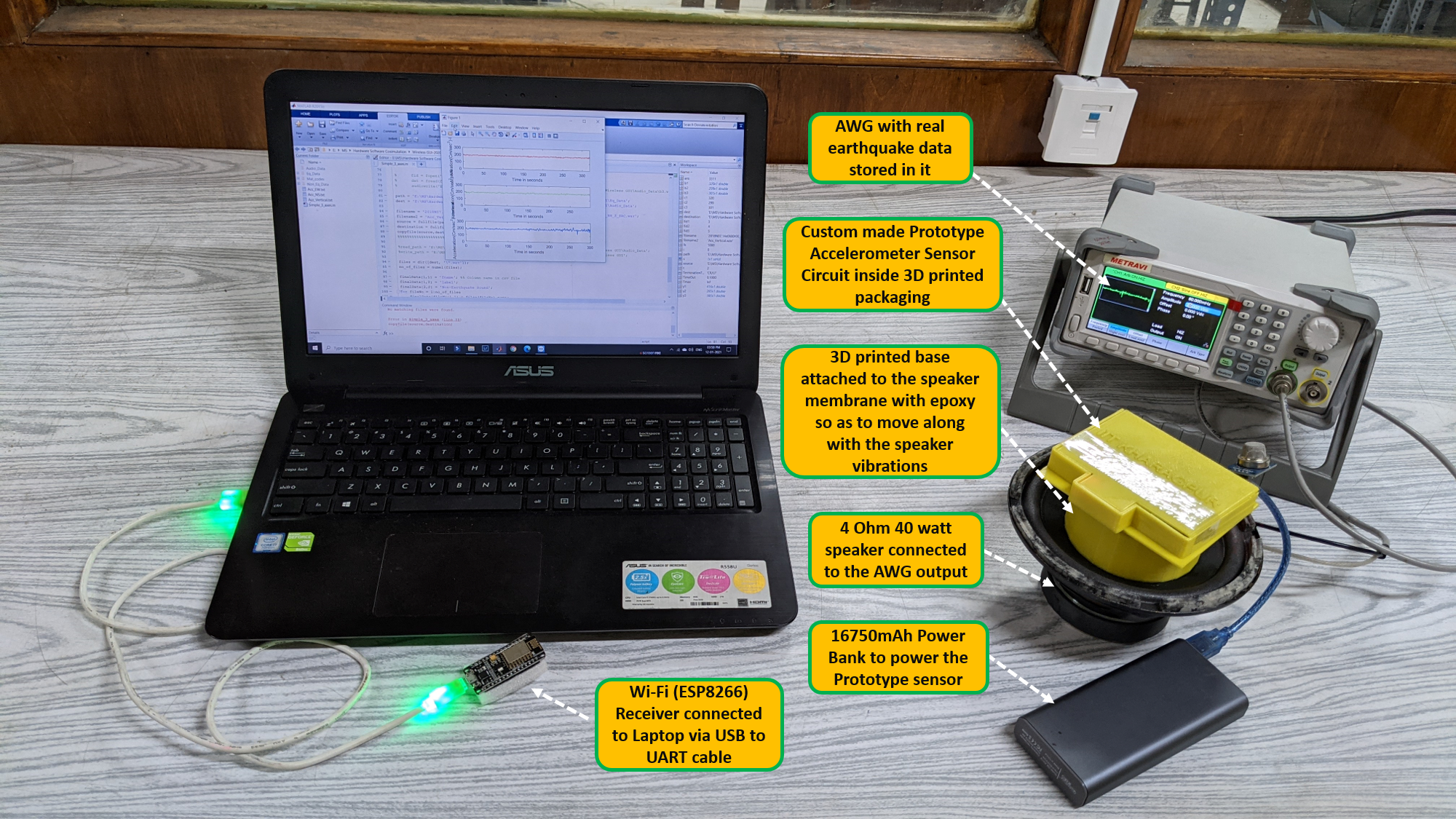}\\
 \caption{Test Setup for the entire System.}
 \label{Figure 14}
  \end{center}
\end{figure}

Figure: \ref{Figure 13(b)} and Figure: \ref{Figure 13(c)} gives us the image of the 3D printed packaging solution used to house the sensor board and the exact specification details of the components used to design the sensor PCB respectively.
After the Deep learning models were validated and finalized, an interesting experimental setup was made so as to test the effectiveness and robustness of the models. The idea was to mimic the vibrations from an actual earthquake signal, sense it with the help of the custom designed sensor PCB, transmit the data wirelessly to a remote computer where the data gets plotted, processed and stored in real time. Also, as soon as the Deep Learning models detects the required data, it activates and predicts the label so as to classify whether or not the recorded data corresponds to an earthquake or a non earthquake sound. The experimental setup can be visualized from Figure: \ref{Figure 14}.

The test setup and the overall system included the following steps:
\begin{itemize}
    \item An Arbitary Waveform Generator (AWG). Real earthquake data was stored in it so that it could generate the output from that data.
    \item A 4 ohm 40 Watt Speaker. The output of the AWG was connected to the speaker so as to make the speaker vibrate according to the output waveform and thus mimic an earthquake.
    \item The Prototype circuit board consisting of the accelerometer sensor to sense the speaker vibrations, microcontroller and an ESP8266 wifi communication block (transmitter built within the PCB and receiver is coonected to the remote PC via a USB to UART cable).
    \item MATLAB code running in the remote PC to receive, process, plot and store the data in real time.
    \item Python code running simultaneously along with the MATLAB code in the remote PC, monitoring the stored data directory to activate the deep learning model for prediction as soon as it detects any data in that directory.
\end{itemize}

The test run of the system was a success as it was able to produce the desired results by accurately predicting the class of the vibration signals from the speaker received by the wireless ESP8266 Wi-Fi module.
\section{Conclusion}
In this paper, a new way of automatic classification of earthquake signals is presented based on CNN and LSTM by using only MFCC features extracted from the waveform. The performance of this algorithm has been tested by its application to regional and local earthquake events selected from PESMOS (Program for Excellence in Strong Motion Studies, IIT Roorkee) and EGMB (Eastern Ghats Mobile Belt, Geology and Geophysics Lab IIT Kharagpur) data sets. Using optimal parameters, for 1000 Hz audio data set the CNN model showed a testing accuracy of 91.102\% for a 0.2-second sample window while the LSTM model obtained an accuracy of 93.999\% for the same. This brings down the standard alarm generation time to approximately 2 seconds after P-wave arrival. The results outperform the conventional algorithm of STA/LTA in terms of classification accuracy with respect to  classification speed and constraint requirements.    
While the models were mainly aimed at classifying earthquake events as quickly as possible, they can also be easily tweaked to give early estimates of magnitudes, ground velocities, shaking intensity, and many more useful parameters. \\
Also, the uniquely innovative experimental setup helped in creating real time earthquake simulations in a cheap, precise and effective way.\\
MFCC for earthquake detection is like putting our ears to hear and listen the sounds inside the earth which we otherwise could not hear and thus treating the earth sounds as earth's speech signals.
The most interesting and effective part of this model is that it can be trained on various classes of sounds other than earthquake sounds alone, to classify every signal that the sensor detects, using their sound signatures. This can be of enormous help in military applications also. If trained with human movement sounds, the model could be deployed in the border and other high-security areas so as to provide us instant information regarding trespassing and other such unlawful activities, releasing the burden to some extent from the security officials and the soldiers. This could be a solution not just for earthquake detection alone but for many other such applications also.

\section*{Acknowledgments}
This work is funded by the Ministry of Human Resource Development(MHRD), Government of India. The authors are thankful to the Department of Earthquake Engineering, IIT Roorkee, and the Department of Geology and Geophysics, IIT Kharagpur, for proving Program for Excellence in Strong Motion Studies(PESMOS) and Easter Ghats Mobile Belt(EGMB) earthquake datasets respectively for the research. This work would not have been possible without the research facilities provided by IIT Kharagpur. Finally, authors are very thankful to all the members of Image Processing and Computer Vision Laboratory, Department of Electronics and Electrical Communication Engineering and Computational Laboratory, Department of Geology and Geophysics, Indian Institute of Technology Kharagpur, for their kind help and support.

\bibliographystyle{unsrt}  
\bibliography{eq}






\end{document}